%% file: main.tex
\documentclass[submission,copyright,creativecommons]{eptcs}
\providecommand{\event}{PxTP 2019} 
\ifdefined\finalizemint\usepackage[finalizecache]{minted}\else\usepackage[frozencache]{minted}\fi

\usepackage{tcolorbox}
\usepackage{etoolbox}
\BeforeBeginEnvironment{minted}{\begin{tcolorbox}}%
\AfterEndEnvironment{minted}{\end{tcolorbox}}%
\usepackage{mdframed}
\usepackage{colortbl}
\usepackage{mdframed}
\usepackage{underscore}           
\usepackage{xspace}
\usepackage{amssymb}
\usepackage{pifont}
\usepackage{comment}
\usepackage{xcolor}
\usepackage{listings}
\usepackage{wrapfig}
\usepackage{graphicx}
\usepackage{semantic}
\usepackage{tikz}
\usepackage{wasysym}
\usepackage{cleveref}

\newcommand\mytitle{Verifying Bit-vector Invertibility Conditions in Coq -- Extended Abstract}

\hypersetup{
 pdfborder={0 0 0},
 colorlinks=true,
 linkcolor=blue,
 urlcolor=blue,
 citecolor=blue,
 pdfauthor={Burak Ekici et al.},
 pdftitle=\mytitle
}

\input{macros.tex}

\title{\mytitle
\thanks{This work has been partially supported by 
the Austrian Science Fund (FWF) grant P26201, 
the European Research Council (ERC) Grant No. 714034 SMART, 
DARPA award N66001-18-C-4012, and 
ONR contract N68335-17-C-0558.
}
}

\author{
Burak Ekici
\institute{University of Innsbruck\\
	Innsbruck, Austria}
\email{burak.ekici@uibk.ac.at}
\and
Arjun Viswanathan
\institute{University of Iowa\\ Iowa City, USA}
\email{arjun-viswanathan@uiowa.edu}
\and
Yoni Zohar
\institute{Stanford University\\ Stanford, USA}
\email{yoniz@cs.stanford.edu}
\and
Clark Barrett
\institute{Stanford University \\ Stanford, USA}
\email{barrett@cs.stanford.edu}
\and
Cesare Tinelli
\institute{University of Iowa\\ Iowa City, USA}
\email{cesare-tinelli@uiowa.edu}
}
\def\titlerunning{Verifying Bit-vector Invertibility Conditions 
in Coq}
\def\authorrunning{B. Ekici, A. Viswanathan, Y. Zohar, C. Barrett, and C. Tinelli}
\begin{document}
\maketitle

\begin{abstract}
This work is a part of an ongoing effort to prove the correctness of
invertibility conditions for the theory of fixed-width bit-vectors, which are
used to solve quantified bit-vector formulas in the Satisfiability Modulo
Theories (\smt) solver CVC4.  While many of these were proved in a completely
automatic fashion for any bit-width, some were only proved for bit-widths up
to 65, even though they are being used to solve formulas over arbitrary
bit-widths.  In this paper we describe our initial efforts in proving a
subset of these invertibility conditions in the \coq proof assistant.  We
describe the \coq library that we use, as well as the extensions that we
introduced to it.
\end{abstract}

\section{Introduction}

Reasoning logically about bit-vectors is useful for many applications in hardware and software verification. 
While Satisfiability Modulo Theories (\smt) solvers 
are able to reason about bit-vectors of fixed width, 
they currently require all widths to be expressed concretely (by a numeral)
in their input formulas.
For this reason, they cannot be used to prove properties of bit-vector operators 
that are parametric in the bit-width
such as, for instance, the associativity of bit-vector concatenation.
Proof assistants such as \coq~\cite{coqref}, that have direct support for
dependent types
are better suited for such tasks.

Bit-vector formulas that are parametric in the bit-width
arise in the verification of parametric Boolean functions 
and circuits (see, e.g.,~\cite{Gupta:1993:RSM:259794.259827}).
In our case, we are mainly interested in parametric lemmas that are relevant 
to internal techniques of SMT solvers for the theory of fixed-width bit-vectors. 
Such techniques are developed a priori for every possible bit-width, 
even though they are applied on a particular bit-width.
Meta-reasoning about the correctness of such solvers then requires 
bit-width independent reasoning.

An example of the latter kind, which is the focus of the current paper,
is the notion of \emph{invertibility conditions}~\cite{DBLP:conf/cav/NiemetzPRBT18}
as a basis for a quantifier-instantiation technique to reason about the satisfiability
of quantified bit-vector formulas.
For a trivial case of an invertibility condition consider the equation
$\equal {\bvadd x s} t$ where 
$x$, $s$ and $t$ are variables of the same bit-vector sort, and
$\bvaddf$ is bit-vector addition.
In the terminology of Niemetz et al.~\cite{DBLP:conf/cav/NiemetzPRBT18}, 
this equation is ``invertible''
for $x$, i.e., solvable for $x$, for any value of $s$ and $t$.
A general solution is represented by the term $\bvsub{t}{s}$.
Since the solution is unconditional, the invertibility condition for $\equal {\bvadd x s} t$
is simply the universally true formula $\true$. 
The formula stating this fact, referred to here as an \emph{invertibility equivalence}, is 
$\true \Leftrightarrow \exists x.\,{\equal {\bvadd x s} t}$,
a valid formula in the theory of fixed-width bit-vectors for any bit-width $n$ 
for $x$, $s$ and $t$.
In contrast, the equation $\equal{\bvmul x s} t$ is not always invertible for $x$ ($\bvmulf$ stands for bit-vector multiplication).
A necessary and sufficient condition for invertibility is
$\equal {\bvand{(\bvor {\bvneg s} s)} t} t$
meaning that the invertibility equivalence
${\equal {\bvand{(\bvor {\bvneg s} s)} t} t} \Leftrightarrow \exists x.\,{\equal{\bvmul x s} t}$
is valid for any bit-width $n$ for $x$, $s$ and $t$~\cite{DBLP:conf/cav/NiemetzPRBT18}.
Notice that this invertibility condition involves the  operations $\bvandf$, $\bvorf$ and $\bvnegf$, and not $\bvmulf$ that occurs in the literal itself.
Niemetz et al.~\cite{DBLP:conf/cav/NiemetzPRBT18} provide a total of
160 invertibility conditions covering several bit-vector operators for both equations
and inequations.
However, they were able to verify, using \smt solvers, 
the corresponding invertibility equivalences only for concrete bit-widths up to 65,
given the reasoning limitations of \smt solvers mentioned earlier. 
A recent paper by Niemetz et al.~\cite{cade2019amaycc} addresses this challenge
by translating these invertibility equivalences
into quantified formulas over the combined theory of non-linear integer arithmetic and 
uninterpreted functions --- a theory supported by a number of \smt solvers.
While partially successful, 
this approach failed to 
verify over a quarter of the invertibility equivalences.

In this work, we approach the task of verifying the invertibility equivalences
proposed in~\cite{DBLP:conf/cav/NiemetzPRBT18}
by proving them interactively with the \coq proof assistant. 
We extend a rich \coq library for bit-vectors we developed in previous work~\cite{DBLP:conf/cav/EkiciMTKKRB17} 
with additional operators and lemmas
to facilitate the task of verifying invertibility equivalences for arbitrary bit-widths, 
and prove a representative subset of them.
Our results offer evidence that proof assistants can support automated theorem provers 
in meta-verification tasks.

Our \coq library models the theory of fixed-width bit-vectors adopted 
by the SMT-LIB 2 standard~\cite{SMTLib2010}.\footnote{
The \smtlib theory is defined at \url{http://www.smt-lib.org/theories.shtml}.
}
It represents bit-vectors as lists of Booleans. The bit-vector type 
is dependent on a positive integer that represents the length of the list. 
Underneath the dependent representation is a simply-typed 
or \emph{raw} bit-vector type
with a size function which is used to explicitly state 
facts on the length of the list. 
A functor translates an instance 
of a raw bit-vector along with specific information 
about its size into a dependently-typed bit-vector.
For this work, we extended the library
with the arithmetic right shift operation and the 
unsigned weak less-than and greater-than predicates 
and proved 
18 invertibility equivalences.
We initially proved these 
equivalences over raw bit-vectors and then 
used these proofs when proving the invertibility 
equivalences over dependent bit-vectors, as we explain
in \Cref{coqbvlib}.

The remainder of this paper is organized as follows.  After some
technical preliminaries in \Cref{sec:prelim}, we provide 
 an overview of invertibility conditions for the
theory of fixed-width bit-vectors in
\Cref{sec:invcond} and discuss previous attempts to
verify them.  Then, in \Cref{coqbvlib}, we describe the bit-vector
\coq library and our current extensions to it.
In \Cref{verifincoq}, we outline how we used the extended library 
to prove the correctness of a representative subset of invertibility
equivalences.  We conclude in \Cref{conc} with directions for
future work.

\section{Preliminaries}
\label{sec:prelim}
We assume the usual terminology of many-sorted first-order logic with equality
(see, e.g.,~\cite{ENDERTON200167} for more details).  We denote equality 
by $=$, and use $x\neq y$ as an abbreviation for $\neg(x=y)$.  The
signature $\sigbv$ of the \smtlib theory of fixed-width bit-vectors
includes a unique sort for each positive integer $n$, which we
denote here by $\sortbv{n}$.  For every positive integer $n$ and a bit-vector of
width $n$, the signature includes a constant of sort $\sortbv{n}$ 
in $\sigbv$ representing that
bit-vector, which we denote as a binary string of length $n$.  The
function and predicate symbols of $\sigbv$ are as described 
in the \smtlib standard.
Formulas of
$\sigbv$ are built from variables (sorted by the sorts $\sortbv{n}$),
bit-vector constants, and the function and predicate symbols of $\sigbv$,
along with the usual logical connectives and quantifiers.  
We write
$\psi[x_{1}\til x_{n}]$ to represent a formula whose free variables are
from the set $\{x_{1}\til x_{n}\}$. 

The semantics of $\sigbv$-formulas is given by interpretations that extend a
single many-sorted first-order structure so that the domain of every sort
$\sortbv{n}$ is the set of bit-vectors of bit-width $n$, and the function and
predicate symbols are interpreted as specified by the \smtlib standard.  A
$\sigbv$-formula is \emph{valid} in the theory of fixed-width bit-vectors
if it evaluates to true in every such interpretation.

In what follows, we denote by $\coqsig$ the sub-signature of $\sigbv$
containing the predicate symbols 
$\bvultf$, $\bvugtf$, $\bvulef$, $\bvugef$ 
(corresponding to strong and weak unsigned comparisons
between bit-vectors, respectively), as
well as the function symbols $\bvaddf$ (bit-vector addition), $\bvandf$, $\bvorf$, $\bvnot$ (bit-wise conjunction,
disjunction and negation), 
$\bvnegf$ (2's complement unary negation), 
and $\bvshl$, $\bvlshrf$ and
$\bvashrf$ (left shift, and logical and 
arithmetical right shifts).  
We also
denote by $\cavsig$ the extension of $\coqsig$ with the predicate symbols
$\bvsltf$, $\bvsgtf$, $\bvslef$, and $\bvsgef$ 
(corresponding to strong and weak signed comparisons
between bit-vectors, respectively), as 
well as the function symbols $\bvsubf$,
$\bvmulf$, $\bvudivf$, $\bvuremf$ (corresponding to subtraction,
multiplication, division and remainder), 
and $\bvconcatf$ (concatenation).  
We use $0$ to represent the bit-vectors composed of all $0$-bits.
Its numerical or bit-vector interpretation should be clear from context.  
Using bit-wise negation $\bvnotf$, we can express the bit-vectors composed
of all $1$-bits by $\bvnot{0}$.

\section{Invertibility Conditions And Their Verification}
\label{sec:invcond}
Many applications rely on bit-precise reasoning and thus can be modeled using the \smtlib theory of fixed-width bit-vectors. 
For certain applications, such as verification of safety properties for programs, 
quantifier-free reasoning is not enough, and the
combination of bit-precise reasoning with the ability to handle quantifiers is needed.
Niemetz et al.~present a technique to solve quantified bit-vector 
formulas, which is based on 
\emph{invertibility conditions}~\cite{DBLP:conf/cav/NiemetzPRBT18}.
An invertibility condition for a variable $x$ in a $\sigbv$-literal $\ell[x,s,t]$ is
  a formula $IC[s,t]$ 
   such that
  $\forall s.\,\forall t.\,IC[s,t] \Leftrightarrow \exists x.\ell[x,s,t]$ is valid in the theory of fixed-width bit-vectors.
For example, consider the bit-vector literal $\bvand{x}{s}\teq t$
where $x$, $s$ and $t$ are distinct variables of the same sort.
The invertibility condition for $x$ given in~\cite{DBLP:conf/cav/NiemetzPRBT18}
is $\bvand{t}{s}\teq{t}$.

Niemetz et al.~\cite{DBLP:conf/cav/NiemetzPRBT18} define invertibility conditions for 
a representative set of literals $\ell$ having a single occurrence of $x$,
that involve the bit-vector operators of $\cavsig$.
The soundness of the technique proposed in that work 
relies on the correctness of the invertibility conditions.
Every literal $\ell[x,s,t]$ and its corresponding invertibility condition $IC[s,t]$
induce the \emph{invertibility equivalence}
\begin{equation}
  IC[s,t]\Leftrightarrow\exists x.\ell[x,s,t]
  \label{IE}
\end{equation}
The correctness of invertibility equivalences should be verified for all
possible sorts for the variables $x,s,t$ for which the condition is well sorted. 
More concretely, for the case where $x,s,t$ are all of sort $\sortbv{n}$, say,
this means that one needs to prove, for \emph{all} $n >0$, the validity of 
\[
\forall s:\sortbv{n}.\,\forall t:\sortbv{n}.\,IC[s,t] \Leftrightarrow \exists x:\sortbv{n}.\ell[x,s,t] \ .
\]
This was done in Niemetz et al.~\cite{DBLP:conf/cav/NiemetzPRBT18}
using an \smt solver but only for concrete values of $n$ from $1$ to $65$.
A proof of \Cref{IE} that is parametric in the bit-width $n$ cannot be done with 
SMT solvers, since they currently only support the theory of \emph{fixed-width} bit-vectors,
where \Cref{IE} cannot even be expressed.
To overcome this limitation, a later paper by Niemetz et al.~\cite{cade2019amaycc} 
suggested a translation from bit-vector formulas with \emph{parametric} bit-widths 
to the theory of (non-linear) integer arithmetic with uninterpreted functions.  
Thanks to this translation, the authors were able to verify, with the aid of \smt solvers
for the theory of integer arithmetic with uninterpreted functions,
the correctness of 110 out of 160 invertibility equivalences.
None of the solvers used in that work were able to prove the remaining equivalences.
For those, it then seems appropriate to use a proof-assistant, 
as this allows for more intervention by the user who can provide crucial intermediate steps.  
It goes without saying that even for the 110 invertibility equivalences that were proved, 
the level of confidence achieved by proving them in a proof-assistant such as \coq 
would be greater than a verification (without a verified formal proof) 
by an \smt solver.  

In the rest of this paper we describe our initial efforts and
future plans for proving the invertibility equivalences, starting with those
that were not proved in~\cite{cade2019amaycc}.

\section{The \coq Bit-vector Library} \label{coqbvlib} 

In this section, we describe the \coq library we use and the extensions
we developed with the goal of formalizing and proving invertibility equivalences.
The original library was developed for \smtcoq~\cite{DBLP:conf/cav/EkiciMTKKRB17}, 
a \coq  plugin that enables \coq to dispatch proofs to external proof-producing solvers. 
It is used to represent \smtlib bit-vectors in \coq.
\coq's own library of bit-vectors~\cite{coqbvlib} was an alternative, but
it has only definitions and no lemmas.  A more suitable substitute could have
been the Bedrock Bit Vectors Library~\cite{bbvlib} or 
the SSRBit Library~\cite{ssrbit}.  We chose the \smtcoq
library mainly because it was explicitly developed
to represent \smtlib bit-vectors in \coq and 
comes with a rich set of lemmas relevant to proving 
the invertibility equivalences.

The \smtcoq library contains both a simply-typed and dependently-typed theory 
of bit-vectors implemented as module types.  The former, which we also refer to
as a theory of \emph{raw bit-vectors}, formalizes bit-vectors as Boolean
lists while the latter defines a bit-vector as a \coq record, with its size as the
parameter, made of two fields: a Boolean list and a coherence condition to
ensure that the parameterized size is indeed the length of the given list. The
library also implements a functor module from the simply-typed module
to the dependently-typed module establishing a correspondence between the two theories.
This way, one can first prove a bit-vector property in the context of the simply-typed theory 
and then map it to its corresponding dependently-typed one via the functor module. 
Note that while it is possible to define bit-vectors natively as a dependently-typed 
theory in \coq and prove their properties there, it would be cumbersome and unduly
complex to do dependent pattern matching or case analysis over bit-vector instances
because of the complications brought by unification in \coq (which is
inherently undecidable).  
One can try to handle such complications as illustrated by Sozeau~\cite{DBLP:conf/itp/Sozeau10}. However, we found the two-theory approach of Ekici et al.~\cite{DBLP:conf/cav/EkiciMTKKRB17} more convenient in practice for our purposes.

The library adopts the little-endian notation for bit-vectors,
thus following the 
internal representation of bit-vectors in SMT solvers such as \cvcfour. 
This makes arithmetic operations 
easier to perform since the least significant bit of a bit-vector is 
the head of the list representing it in the \emph{raw} theory.

Out of the 11 bit-vector operators and 10 predicates 
contained in $\cavsig$,
the library had support 
for 8 operators and 6 predicates.
The supported predicates, however, can be used to express the other 4.
The predicate and function symbols that were not directly supported by the 
library were the weak inequalities 
\bvulef, \bvugef, \bvslef, \bvsgef and the operators
\bvashrf, \bvudivf, and \bvuremf. 
We extended the library with the operator \bvashrf and the predicates \bvulef and \bvugef
and redefined \bvshlf and 
\bvlshrf, as explained in \Cref{verifincoq}. 

We focused on invertibility conditions for 
literals of the form
$x\op s \rel t$ and $s\op x\rel t$,
where $x$, $s$ and $t$ are variables and $\op$ and $\rel$ are respectively function and predicate symbols in $\coqsig\cup\set{=,\neq}$ 
(invertibility conditions for such literals were found in \cite{DBLP:conf/cav/NiemetzPRBT18} for the extended signature $\cavsig$).
$\coqsig$ was chosen as a representative set because it seemed both expressive
enough and feasible for proofs in \coq.
Such literals, as well as their invertibility conditions, include only
operators that are supported by the library (after its extension with
$\bvashrf$, $\bvulef$, and $\bvugef$).

\begin{figure}[t]
	\centering
\begin{minted}[fontsize=\footnotesize,xleftmargin=1em,linenos=true, escapeinside=!!]{coq}
Fixpoint ule_list_big_endian (x y : list bool) :=
  match x, y with
  | nil, nil => true
  | nil, _ => false 
  | _, nil => false 
  | xi :: x', yi :: y' => ((eqb xi yi) && (ule_list_big_endian x' y')) 
                           || ((negb xi) && yi)
  end. 

Definition ule_list (x y: list bool) :=
  (ule_list_big_endian (rev x) (rev y)).

Definition bv_ule (a b : bitvector) :=
  if @size a =? @size b then 
    ule_list a b 
  else 
    false.
	\end{minted}
  \caption{Definitions of $\bvulef$ in \coq.}
	\label{fig:bvule}
\end{figure}

To demonstrate the intuition and various
aspects of the extension of the library, we briefly
describe the addition of \bvulef\ (the definition 
of \bvugef is similar).
The relevant \coq definitions are provided in 
\Cref{fig:bvule}.\footnote{Both the library and the proofs of invertibility equivalences can be found at \url{https://github.com/ekiciburak/bitvector/tree/pxtp2019}. It compiles with \texttt{coqc-8.9.0}.}  
Like most other operators, \bvulef is defined in several \emph{layers}.
The function \texttt{bv\_ule}, at the highest layer, 
ensures that comparisons are 
between bit-vectors of the same size and then calls
\texttt{ule\_list}.  
Since we want to compare bit-vectors starting from their 
most significant bits and the input lists start instead
with the least significant bits (because of the little-endian encoding),
\texttt{ule\_list} first reverses the two lists.
Then it calls \texttt{ule\_list\_big\_endian}, which 
we consider to be at the lowest layer of the definition.
\texttt{ule\_list\_big\_endian} then does a lexicographical 
comparison of the two lists, starting from 
the most significant bits.

To see why the addition of \bvulef to the library is useful, consider, for example, the following parametric lemma, stating that $\bvnot{0}$ is the largest unsigned bit-vector of its type:
\begin{equation}
  \forall x:\sortbv{n}.\,\bvule{x}{\bvnot{0}}
	\label{eq:ule1}
\end{equation}
 When not using this explicit operator, we usually rewrite it as:
\begin{equation}
  \forall x:\sortbv{n}.\,\bvult{x}{\bvnot{0}}\lor{\equal{x}{\bvnot{0}}}
\end{equation}
In such cases, since the definitions of $\bvultf$ and $=$ have a similar structure to the one in \Cref{fig:bvule},  
we 
 strip down the layers of 
\bvultf and \teq separately, whereas using $\bvulef$, we only do this once. 
Depending on the specific proof at hand, using $\bvulef$ is sometimes more convenient for this reason.

\section{Proving Invertibility Equivalences in \coq}
\label{verifincoq}
In this section we provide specific details about proving invertibility equivalences in \coq.
In addition to the bit-vector library described in \Cref{coqbvlib},
in several proofs of invertibility equivalences we benefited from 
\coqhammer~\cite{DBLP:journals/jar/CzajkaK18}, a plug-in that aims at extending the 
automation in \coq by combining machine learning and automated reasoning
techniques in a similar fashion to what is done in Isabelle/HOL~\cite{nipkow2002isabelle}.
Note that one does
not need to install \coqhammer in order to build the bit-vector library, 
since all the proof reconstruction
tactics of \coqhammer are included in it.

The natural representation of bit-vectors in \coq 
is the dependently-typed representation,
and therefore the invertibility equivalences are formulated using this representation.
As discussed in \Cref{coqbvlib}, however, proofs in this representation  
are composed of proofs over simply-typed
bit-vectors, which are easier to reason about.  
Some conversions between the different representations are then needed
to lift a proof over raw bit-vectors to one over dependently-typed bit-vectors.
%
%

For example, \Cref{fig:dep-proof} includes a proof
of the following direction of the invertibility equivalence for \bvashrf and \bvultf:
\begin{equation}
	\forall s:\sortbv{n}.\,\forall t:\sortbv{n}.\, 
	 \imp{(\exists x:\sortbv{n}.
	 \bvult{\bvashr{s}{x}}{t})}{(\booland{(\boolor{\bvult{s}{t}}
	 {\boolnot(\bvslt{s}{0})})}{t \neq 0})}
\end{equation}
In the proof, lines \ref{firsttrans}--\ref{lasttrans} transform
the dependent bit-vectors from the goal and the hypotheses 
into simply-typed bit-vectors. Then, lines
\ref{simplproof}--\ref{lastline} invoke 
the corresponding lemma for  
simply-typed bit-vectors (called \texttt{InvCond.bvashr\_ult2\_rtl}) 
along with some simplifications.

\begin{figure}[t]
  \centering
    \begin{minted}[fontsize=\footnotesize,xleftmargin=1em,linenos=true, escapeinside=!!]{coq}
Theorem bvashr_ult2_rtl : forall (n : N), forall (s t : bitvector n),
(exists (x : bitvector n), (bv_ult (bv_ashr_a s x) t = true)) ->
(((bv_ult s t = true) \/ (bv_slt s (zeros n)) = false) /\ 
(bv_eq t (zeros n)) = false).
Proof. intros n s t H. 
	destruct H as ((x, Hx), H).!\label{firsttrans}!
	destruct s as (s, Hs).
	destruct t as (t, Ht).
	unfold bv_ult, bv_slt, bv_ashr_a, bv_eq, bv in *. cbn in *.!\label{lasttrans}!
	specialize (InvCond.bvashr_ult2_rtl n s t Hs Ht); intro STIC.!\label{simplproof}!
	rewrite Hs, Ht in STIC. apply STIC.
	now exists x. !\label{lastline}!
Qed.
\end{minted}
  \caption{A proof of one direction of the invertibility equivalence for
    \bvashrf and \bvultf using dependent types.}
\label{fig:dep-proof}
\end{figure}

Most of the effort in this project went into proving 
equivalences over raw bit-vectors. 
As an illustration, 
consider the following equivalence 
over \bvshlf and \bvugtf:
\begin{equation}
  \forall s:\sortbv{n}.\,\forall t:\sortbv{n}.\, 
  (\bvult{t}{\bvshl{\bvnot{0}}{s}})
\Leftrightarrow
  (\exists x:\sortbv{n}.\bvugt{\bvshl{x}{s}}{t})
  \label{eq:iccoqproof}
\end{equation}
The left-to-right implication is easy to prove 
using $\bvnot{0}$ itself as the witness of the existential
proof goal 
and
considering the symmetry between $\bvugtf$ and $\bvultf$.
The proof of the right-to-left implication 
relies on the following lemma:
\begin{equation}  
  \forall x:\sortbv{n}.\,\forall s:\sortbv{n}.\,\bvule{(\bvshl{x}{s})}{(\bvshl{\bvnot{0}}{s})}
  \label{eq:ltxs1s}
\end{equation}
From the right side of the equivalence in \Cref{eq:iccoqproof}, we get
some $x$ for which $\bvugt{\bvshl{x}{s}}{t}$ holds.
Flipping the inequality, we have that 
$\bvult{t}{\bvshl{x}{s}}$; using this, and transitivity over 
\bvultf and \bvulef, 
Lemma~\ref{eq:ltxs1s} gives us 
the left side of the equivalence in \Cref{eq:iccoqproof}.

As mentioned in \Cref{coqbvlib},
we have redefined the shift operators $\bvshlf$ and $\bvlshrf$ in the library. This was 
instrumental, for example, in the proof of \Cref{eq:ltxs1s}.
\Cref{fig:bvshl-def} 
includes both the original and new definitions of \bvshlf. 
The definitions 
of \bvlshrf are similar.
Originally,
$\bvshlf$
was defined using the
\texttt{shl\_one\_bit} and the \texttt{shl\_n\_bits}
functions. 
\texttt{shl\_one\_bit} shifts 
the bit-vector to the left by one bit and is repeatedly called by \texttt{shl\_n\_bits}
to complete the shift.
The new definition \texttt{shl\_n\_bits\_a} uses 
\texttt{mk\_list\_false} which constructs the necessary list of $0$s and
appends (\texttt{++} in \coq) it to the beginning of the list (because of
the little-endian encoding); 
the bits to be shifted from the original
bit-vector are retrieved using the \texttt{firstn} function,
which is defined in the \coq library for lists.
The \texttt{nat}
type used in \Cref{fig:bvshl-def} is the \coq representation of Peano natural
numbers that has $\texttt{0}$ and $\texttt{S}$ as its two constructors 
--- as depicted in the pattern match in lines
\ref{zero} and \ref{succ}. 
The theorem at the bottom of \Cref{fig:bvshl-def}
allows us to switch between the two definitions when needed.
Function \texttt{bv\_shl} defines the left shift operation using 
\texttt{shl\_n\_bits} whereas \texttt{bv\_shl\_a} does it 
using \texttt{shl\_n\_bits\_a}.

\begin{figure}[t]
  \centering
    \begin{minted}[fontsize=\footnotesize,xleftmargin=1em,linenos=true, escapeinside=!!]{coq}
Definition shl_one_bit  (a: list bool) :=
  match a with
	| [] => []
	| _ => false :: removelast a 
  end.

Fixpoint shl_n_bits  (a: list bool) (n: nat) :=
  match n with
	| O => a !\label{zero}!
	| S n' => shl_n_bits (shl_one_bit a) n'  !\label{succ}!
  end.

Definition shl_n_bits_a  (a: list bool) (n: nat) :=
  if (n <? length a)%nat then !\label{natlt}!
	mk_list_false n ++ firstn (length a - n) a
  else 
	mk_list_false (length a).

Theorem bv_shl_eq: forall (a b : bitvector), bv_shl a b = bv_shl_a a b.
\end{minted}
  \caption{Various definitions of \bvshlf.}
\label{fig:bvshl-def}
\end{figure}

The new definition 
uses \texttt{firstn} and 
\texttt{++}, over which many necessary properties 
are already proven in the standard library. This 
benefits us in manual proofs, and in calls to 
\coqhammer,
since the latter is able to use lemmas from the imported libraries
to prove the goals that are given to it.
Using this representation, proving  
\Cref{eq:ltxs1s} reduces to proving Lemmas
\texttt{bv\_ule\_1\_firstn} and
\texttt{bv\_ule\_pre\_append}, shown in 
\Cref{fig:lt1-lemmas}. 
The proof of
\texttt{bv\_ule\_pre\_append} benefited from 
the property \texttt{app\_comm\_cons} from the 
standard list library of \coq, while
\texttt{firstn\_length\_le} was useful in reducing 
the goal of \texttt{bv\_ule\_1\_firstn}
to \coq's equivalent of \Cref{eq:ule1}. 
The statements of the properties 
mentioned from the standard library are also 
shown in \Cref{fig:lt1-lemmas}.
\texttt{mk\_list\_true}
creates a bit-vector that represents 
\bvnot{0}, of the length given to it as input, 
and \texttt{bv\_ule} is the representation 
of \bvulef in the bit-vector library.
\texttt{bv\_ule} has output type \texttt{bool} 
(and so we equate terms in which it occurs to \texttt{true}), 
while the functions from the standard library 
have output type \texttt{Prop}.
We also have two definitions for \bvashrf, and a proof
of their equivalence (as done for the other shift operators).

\begin{figure}[t]
	\centering
	\begin{minted}[fontsize=\footnotesize,xleftmargin=1em,linenos=true, escapeinside=!!]{coq}
Lemma bv_ule_1_firstn : forall (n : nat) (x : bitvector), 
	(n < length x)%nat ->
	bv_ule (firstn n x) (firstn n (mk_list_true (length x))) = true.

Lemma bv_ule_pre_append : forall (x y z : bitvector), bv_ule x y  = true ->
	bv_ule (z ++ x) (z ++ y) = true.

Theorem app_comm_cons : forall (x y:list A) (a:A), a :: (x ++ y) = (a :: x) ++ y.

Lemma firstn_length_le: forall l:list A, forall n:nat,
	n <= length l -> length (firstn n l) = n.
	\end{minted}
  \caption{Examples of lemmas used in proofs of invertibility equivalences.}
	\label{fig:lt1-lemmas}
\end{figure}	
%

\Cref{icresults} summarizes the results of proving invertibility equivalences
for invertibility conditions in the signature $\coqsig$.
In the table, 
\coqp means that the invertibility equivalence was successfully verified in \coq
but not in~\cite{cade2019amaycc}, while \cadep means the opposite;
{\both} means that the invertibility equivalence was verified using both approaches, 
and \none means that it was verified with neither.
We successfully proved all invertibility equivalences over $\teq$ that
are expressible in $\coqsig$, including 4 that were not proved in~\cite{cade2019amaycc}. 
For the rest of the predicates, we focused only
on the 8 invertibility equivalences that were not proved in~\cite{cade2019amaycc}, 
and succeeded in proving 7 of them.
Overall, these results strictly improve the results of~\cite{cade2019amaycc},
as we were able to prove 11 additional invertibility equivalences in \coq.
Taking into account our work together with~\cite{cade2019amaycc},
only one invertibility equivalence for the restricted signature 
is not fully proved yet, the one for the literal 
$\bvugt{\bvlshr{x}{s}}{t}$,
although one direction of the equivalence,
namely $IC[s,t]\Rightarrow \exists x.\ell[x,s,t]$,
was successfully proved both in \coq and in~\cite{cade2019amaycc}.

\begin{table}
\begin{center}
{%
  \renewcommand{\arraystretch}{1.2}%
  \begin{tabular}{r@{\hspace{2.0em}}c@{\hspace{1.0em}}c@{\hspace{1.5em}}c@{\hspace{1.0em}}c@{\hspace{1.5em}}c@{\hspace{1.0em}}c}
    \hline
    \\[-2.5ex]
    $\ell[x]$ & \teq & \tneq & \bvultf & \bvugtf & \bvulef &
    \bvugef
    \\[.5ex]
    \hline
    \\[-2.5ex]
    $\bvneg{x}  \rel t$ & \both & \cadep & \cadep & \cadep  
     & \cadep & \cadep \\
    $\bvnot{x}  \rel t$ & \both & \cadep & \cadep & \cadep  
     & \cadep & \cadep  \\
    $\bvand{x}{s}  \rel t$ & \coqp & \cadep & \cadep & \cadep  
     & \cadep & \cadep \\
    $\bvor{x}{s}   \rel t$ & \coqp & \cadep & \cadep & \cadep 
     & \cadep & \cadep \\
    $\bvshl{x}{s}  \rel t$ & \coqp & \coqp & \cadep & \coqp   
     & \cadep & \coqp \\
    $\bvshl{s}{x}  \rel t$ & \both & \cadep & \cadep & \cadep 
     & \cadep & \cadep \\
    $\bvlshr{x}{s} \rel t$ & \both & \cadep & \cadep & \none 
     & \cadep & \cadep \\
    $\bvlshr{s}{x} \rel t$ & \both & \cadep & \cadep & \cadep 
     & \cadep & \cadep \\
    $\bvashr{x}{s} \rel t$ & \coqp & \cadep & \cadep & \cadep 
     & \cadep & \cadep \\
    $\bvashr{s}{x} \rel t$ & \both & \cadep & \coqp & \coqp  
     & \coqp & \coqp \\
    $\bvadd{x}{s}  \rel t$ & \both & \cadep & \cadep & \cadep 
     & \cadep & \cadep \\
  \end{tabular}%
}
\end{center}
  \caption{Proved invertibility equivalences in $\coqsig$ where $\rel$ ranges over the given predicate symbols.
}\label{icresults} 
  \end{table}

\section{Conclusion and Future Work}
\label{conc}
We have described our work-in-progress on verifying bit-vector invertibility conditions
in the \coq proof assistant, which required extending
a bit-vector library in \coq.
The most immediate direction for future work is 
proving more of the invertibility equivalences supported by the bit-vector library.
In addition, we plan to extend the library so that it supports the full syntax in which invertibility conditions
are expressed, namely $\cavsig$. 
We expect this to be useful also for verifying properties about bit-vectors in other applications.

\bibliographystyle{eptcs}
\bibliography{generic}
\end{document}

%% file: macros.tex
\newcommand*{\ttfamilywithbold}{\fontfamily{lmtt}\selectfont}
\newsavebox{\proofbox}

\newcommand{\xmark}{\ding{53}}%

\newcommand{\coqp}{{\Large{\color{black}{$\checkmark$}}\xspace}}
\newcommand{\cadep}{{\color{gray}{$\checkmark$}}\xspace}
\newcommand{\both}{{\coqp\nolinebreak\kern-0.7em\cadep}}
\newcommand{\none}{{\color{black}\xmark}\xspace}

\newcommand{\coq}{Coq\xspace}
\newcommand{\cavsig}{\Sigma_{1}}
\newcommand{\coqsig}{\Sigma_{0}}

\newcommand{\set}[1]{\left\{#1\right\}}
\newcommand{\smtcoq}{SMTCoq\xspace}
\newcommand{\smtlib}{SMT-LIB~2\xspace}
\newcommand{\coqhammer}{CoqHammer\xspace}
\newcommand{\cvcfour}{CVC4\xspace}
\newcommand{\smt}{SMT\xspace}

\newcommand{\op}{\ensuremath{\diamond}\xspace}
\newcommand{\rel}{\ensuremath{\bowtie}\xspace}

\newcommand{\teq}{\ensuremath{=}\xspace}
\newcommand{\tneq}{\ensuremath{\not\teq}\xspace}

\newcommand{\true}{\ensuremath{\top}\xspace}

\newcommand{\til}{,\ldots,}
\newcommand{\sig}{\ensuremath{\Sigma}\xspace}
\newcommand{\sigbv}{\ensuremath{\sig_{BV}}\xspace}
\newcommand{\sort}{\ensuremath{\sigma}\xspace}
\newcommand{\sortbv}[1]{\ensuremath{\sort_{[#1]}}\xspace}
\newcommand{\bvaddf}{\ensuremath{+}\xspace}
\newcommand{\bvsubf}{\ensuremath{-}\xspace}
\newcommand{\bvandf}{\ensuremath{\mathrel{\&}}\xspace}
\newcommand{\bvashrf}{\ensuremath{\mathop{>\kern-.3em>_a}}\xspace}
\newcommand{\bvconcatf}{\ensuremath{\circ}\xspace}
\newcommand{\bvlshrf}{\ensuremath{\mathop{>\kern-.3em>}}\xspace}
\newcommand{\bvmulf}{\ensuremath{\cdot}\xspace}
\newcommand{\bvorf}{\ensuremath{\mid}\xspace}
\newcommand{\bvsgef}{\ensuremath{\ge_s}\xspace}
\newcommand{\bvsgtf}{\ensuremath{>_s}\xspace}
\newcommand{\bvshlf}{\ensuremath{\mathop{<\kern-.3em<}}\xspace}
\newcommand{\bvslef}{\ensuremath{\le_s}\xspace}
\newcommand{\bvsltf}{\ensuremath{<_s}\xspace}
\newcommand{\bvudivf}{\ensuremath{\div}\xspace}
\newcommand{\bvugef}{\ensuremath{\geq_u}\xspace}
\newcommand{\bvugtf}{\ensuremath{>_u}\xspace}
\newcommand{\bvultf}{\ensuremath{<_u}\xspace}
\newcommand{\bvulef}{\ensuremath{\leq_u}\xspace}
\newcommand{\bvuremf}{\ensuremath{\bmod}\xspace}
\newcommand{\bvnegf}{\ensuremath{-}\xspace}
\newcommand{\bvnotf}{\ensuremath{{\sim}\,}\xspace}

\newcommand{\booland}[2]{\ensuremath{#1\,\wedge\,#2}\xspace}
\newcommand{\boolnot}[1]{\ensuremath{\neg #1}\xspace}
\newcommand{\boolor}[2]{\ensuremath{#1\,\vee\,#2}\xspace}
\newcommand{\bvadd}[2]{\ensuremath{#1 \bvaddf #2}\xspace}
\newcommand{\bvand}[2]{\ensuremath{#1 \bvandf #2}\xspace}
\newcommand{\bvashr}[2]{\ensuremath{#1 \bvashrf #2}\xspace}

\newcommand{\bvlshr}[2]{\ensuremath{#1 \bvlshrf #2}\xspace}
\newcommand{\bvmul}[2]{\ensuremath{#1 \bvmulf #2}\xspace}
\newcommand{\bvneg}[1]{\ensuremath{\bvnegf#1}\xspace}
\newcommand{\bvnot}[1]{\ensuremath{\bvnotf\!#1}\xspace}
\newcommand{\bvor}[2]{\ensuremath{#1 \bvorf #2}\xspace}

\newcommand{\bvshl}[2]{\ensuremath{#1 \bvshlf #2}\xspace}

\newcommand{\bvslt}[2]{\ensuremath{#1 \bvsltf #2}\xspace}
\newcommand{\bvsub}[2]{\ensuremath{#1 \bvsubf #2}\xspace}

\newcommand{\bvugt}[2]{\ensuremath{#1 \bvugtf #2}\xspace}
\newcommand{\bvule}[2]{\ensuremath{#1 \bvulef #2}\xspace}
\newcommand{\bvult}[2]{\ensuremath{#1 \bvultf #2}\xspace}

\newcommand{\equal}[2]{\ensuremath{#1 \teq #2}\xspace}
\newcommand{\imp}[2]{\ensuremath{#1\hspace{0.1em}\Rightarrow\hspace{0.1em}#2}\xspace}